\documentclass[journal=apchd5,manuscript=article]{achemso}

\usepackage{chemformula} 
\usepackage[T1]{fontenc} 
\usepackage{ulem}

\author{Alexei V. Prokhorov}
\email{avprokhorov33@mail.ru}
\affiliation[VlSU]
{Department of Physics and Applied Mathematics, Vladimir State University named after A. G. and N. G. Stoletovs (VlSU), Vladimir, Russian}
\alsoaffiliation[MIPT]
{Center for Photonics and 2D Materials, Moscow Institute of Physics and Technology (MIPT), Dolgoprudny, Russian}
\author{Alexander V. Shesterikov}
\affiliation{Department of Physics and Applied Mathematics, Vladimir State University named after A. G. and N. G. Stoletovs (VlSU), Vladimir, Russian}
\alsoaffiliation[MIPT]
{Center for Photonics and 2D Materials, Moscow Institute of Physics and Technology (MIPT), Dolgoprudny, Russian}
\author{Mikhail Yu. Gubin}
\affiliation[VlSU]
{Department of Physics and Applied Mathematics, Vladimir State University named after A. G. and N. G. Stoletovs (VlSU), Vladimir, Russian}
\alsoaffiliation[MIPT]
{Center for Photonics and 2D Materials, Moscow Institute of Physics and Technology (MIPT), Dolgoprudny, Russian}
\author{Valentyn S. Volkov}
\affiliation[MIPT]
{Center for Photonics and 2D Materials, Moscow Institute of Physics and Technology (MIPT), Dolgoprudny, Russian}
\author{Andrey B. Evlyukhin}
\affiliation[LUH]
{Institute of Quantum Optics, Leibniz Universit\"{a}t Hannover (LUH), Hannover, Germany}
\alsoaffiliation[MIPT]
{Center for Photonics and 2D Materials, Moscow Institute of Physics and Technology (MIPT), Dolgoprudny, Russian}
\email{a.b.evlyukhin@daad-alumni.de}
\title{
 {{Quasi-trapped} modes  in metasurfaces of anisotropic MoS$_2$ nanoparticles for  absorption {and polarization} control in the telecom optical range}}
\keywords{anisotropic materials, bianisotropic nanoparticles, all-dielectric metasurface, quasi-trapped mode, collective absorbing, polarization transformation}

\begin{document}

\begin{tocentry}

\includegraphics[width=\columnwidth]{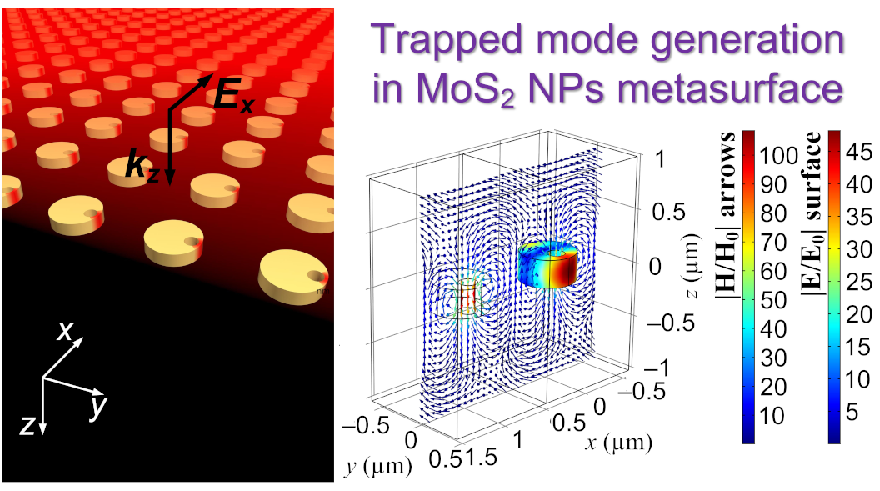}





\end{tocentry}

\begin{abstract}
  We investigate the resonant optical response of single material-anisotropic nanoparticles (NPs) of molybdenum disulfide (MoS$_2$) and  their two-dimensional arrays (metasurfaces) irradiated by  plane waves of the telecomunication optical range. Nanoparticles in the form of a  disk with {centered and shifted} hole are considered. Using the recently experimental measured  the MoS$_2$ dielectric permittivity and numerical calculations with analytical multipole analysis, we show that the material-anisotropy of  NPs can lead to specific  nonlocal contributions  in their magnetic and electric dipole response and affect the effective dipole polarizabilities. Applying a special procedure we determine the period of  the MoS$_2$ metasurfaces supporting the quasi-trapped mode (QTM)  resonance around  the tellecom wavelength of 1550 nm. It is shown that regardless extremely weak  absorption of the single nanoparticles,  the excitation of the QTM leads to effective narrowband absorption in the metasurfaces. {Influences of the  linear polarization direction of normally incident waves  on the QTM implementation and the reflection and transmission spectra are studied.} It is found and demonstrated,  for the first time, that a metasurface, composed of the MoS$_2$ disks with their anisotropy perpendicular to the metasurface plane, has the properties of a continuous birefringent medium. Due to these properties, normally incident and linear-polarized waves can be transformed in the transmitted and reflected waves with {changed} polarizations.
\end{abstract}
\section{Introduction}
Currently, the design of the ultra-narrowband absorbers is of great practical interest. Until recently, the structures and metasurfaces based on the conductive materials have been used to create such absorbers. The configuration and lattice resonances in them allow narrowing the spectral transmission window, while the conductor provides a fast damping of light in it~\cite{Pendry,Liu,ABS1,ABS2}. The composite metal-dielectric absorbers utilize the resonances of a dielectric metamaterial in order to concentrate a light and direct it to a plasmonic substrate for conversion of near field energy to Joule losses~\cite{Kuznetsov}. The quality factor of the considered resonances as the ratio of the resonant wavelength to the full width at half maximum (FWHM) depends on the topology of the structure. For example, in Ref.~\citenum{ABS1} using gold mushrooms on a complex substrate, the quality factor is about 130 in the near-infrared range; in the theoretical work~\cite{ABS2} for an array of gold nanoribbons on a thin gold layer, it reaches 350 in the visible range. In addition, the terahertz absorbers based on the structured graphene~\cite{Yan} with quality factor of about 9 and with quality factor of about several hundred units~\cite{Wu} for a wavelength 1550 nm were proposed. An additional remarkable feature of narrowband absorbers is high sensitivity to the local environment, which allows to detect individual particles and nanoobjects with high precision~\cite{Stebunov,ABS3,ABS4,ABS5}. However, the obligatory presence of a plasmon substrate in such structures allows to use them only as individual devices of limited size. In the case of creating the spatially distributed sensor of a large size or depositing the absorbing coating on a flexible substrate of arbitrary shape, the presented approach is not applicable. Partial solution can be the use of {nanoparticle (NP)}  lattice without plasmonic substrate and the utilizing absorption properties of the NPs forming the lattice. In this case the diffractive-coupled lattice of metal or semiconductor  NPs  provides the  collective lattice resonances with the quality factors determined by the configuration and size perimeters of the system~\cite{evlyukhin2010optical,offermans2011,Evlyukhin_PRB_2012,babicheva2017resonant,kravets2018plasmonic,cherqui2019plasmonic}.  

The recent researches have focused on the development of all-dielectric absorbers and sensors, for which a high level of absorption is achieved by matching impedances and combining electric and magnetic dipole resonances at the same frequency in a metamaterial fabricated from weakly absorbing {NPs}, for example, solid Si {NPs}~\cite{OSA1,babicheva2018}. However, the quality factor of such systems is low and does not exceed 12. Not very high values of quality factor are reached in metasurfaces composed of doped Si building blocks with increased material losses, see Refs.~\citenum{ACS1,AIP1,LPR1}.

The improvement of quality factor control of the all-dielectric absorbers and enhancing sensitivity of the corresponding sensors can be reached, on the one hand, applying new physical mechanism, and, on the other hand, using new optical materials for their fabrication. One such an approach is to use the combination of the fundamental effects of nonlocality  in single NPs optical response and the collective resonances~\cite{ourLPR2020} in all-dielectric metasurfaces  composed of such building blocks~\cite{zhang2013near,tuz2018all}. In particular, the metasurface composed of optically bianisotropic NPs~\cite{B1} can support the quasi-trapped modes {(QTMs)} or, by other words, the quasi bound states in the continuum (q-BICs)~\cite{Koshelev2019}. In general the trapped modes or  BICs are protected eigenmodes of ideal lossless optical systems remaining perfectly localized without radiation into free space~\cite{Krasnok2019} that provides their infinite Q-factor resonances and perfect confinement of optical energy~\cite{Rivas2019,Xiao2019,Rybin2021}. In practical cases, access to similar states can be obtained by weak distortion or perturbation of ideal structure's symmetrical properties that converts the trapped modes into QTMs (q-BICs) \cite{fedotov2007sharp,B4}. For metasurfaces such perturbations can be realized {via} exploiting of especially initiated bianisotropic properties of their building blocks.

Recently, it was analytically demonstrated in Ref.~\citenum{B2} that, in the metasurfaces composed of dielectric nanodisks with an eccentric through hole, the QTMs  are generated due to the self-synchronization for the single disk's magnetic dipole moments oriented orthogonally to the metasurface and excited due to the disk's bianisotropic response. The strength of this effect depends on the values of the nondiagonal elements of the disk's dipole polarizability tensor that is responsible for the bianisotropy~\cite{B1,B2}. In the case of the QTM excitation, the transmission and reflection spectra of arrays of such particles have the ultra-narrow resonances~\cite{B2,B4} which can be associated with the nonlinear~\cite{Miroshnichenko_AdvSci_2019} and absorption (thermal)~\cite{Miroshnichenko_Small_2019} processes.

So far the referenced results related with QTMs concerned the metasurfaces composed of nanoparticles with isotropic dielectric properties. In these cases, for realization of the required resonances of single particles only shape and size factors can be applied. However, the use of 2D layered anisotropic materials~\cite{Novoselov1} such as \ch{MoS2}~\cite{NatPhotVolNov,Chhowalla,Schonfeld,Ermolaev,Yoffe} for metasurface's building blocks can provides an additional degree of freedom for tuning and control of  their optical  resonances. Due to the considerable difference between the refractive indices along and orthogonal to the material layers, their resonant response can be significantly sensitive to the propagating and polarization properties of incident light waves. This is of fundamental importance, since the higher dielectric anisotropy of single nanoparticles in metasurfaces can result in new possibilities with respect to the excitation and manipulations of the QTMs leading to the selective resonant light absorption, transmission, or reflection. In order to clarify and demonstrate these new facilities,  in this article we investigate the features of the trapped mode realization in the metasurfaces composed of \ch{MoS2} disk particles. We show that, due to the QTM response of such metasurfaces, an effective narrow absorption band can be realized in the telecom (wavelength $\sim 1550$~nm) spectral range, where \ch{MoS2} material has very weak anisotropic absorption. {Besides, the influence} of the {polarization rotation  of the incident field, relative to the orientation of \ch{MoS2} layers in the disks,} on polarization changes of transmitted and reflected waves is investigated. 

\section{Resonant absorption in MoS$_2$  metasurfaces}
\begin{figure}[t]
\centering
\includegraphics[width=0.7\columnwidth]{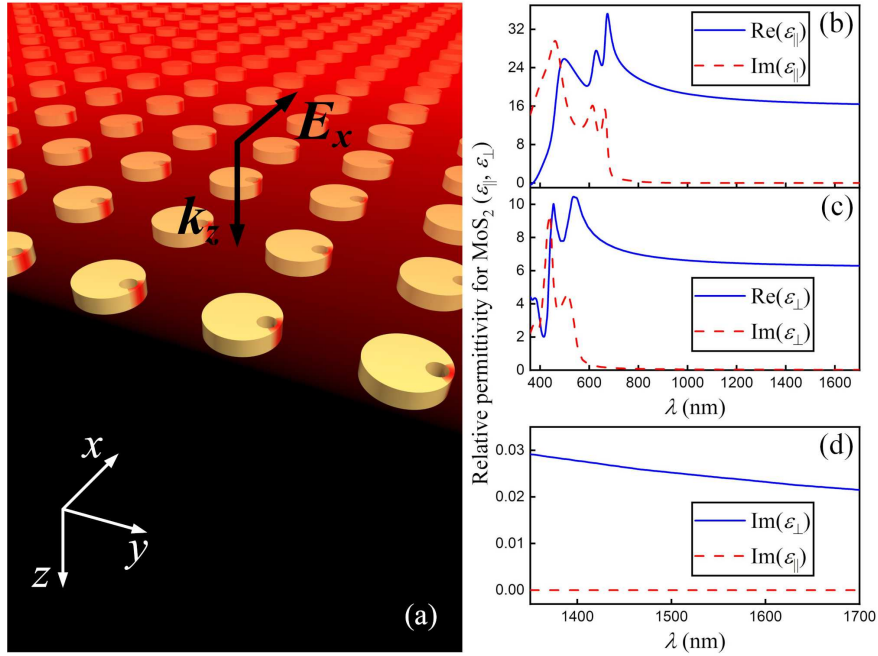}
\caption{\label{fig:1} (a) The model of metasurface composed of hollow \ch{MoS2} disks with visualization of the electric field component in the regime of quasi-trapped mode generation stylized on the base of numerical simulation. Relative dielectric permittivity for MoS$_2$ (b) along and (c) orthogonal to the material layers~\cite{NatPhotVolNov}. {(d) The comparison of imaginary parts of permittivities for mutually orthogonal directions.}}
\end{figure}

In this article, we will consider metasurfaces composed of disk nanoparticles made of  {MoS$_2$} with layers perpendicular to the disk base and parallel to the $ yOz $ plane (Figure~\ref{fig:1}a).  Every disk has radius $R_{2}$, hieght $H$ and a through hole with radius $R_{1}$. The hole can be  shifted on $\Delta$ along $y$-axis, as it is shown in Figure~\ref{fig:1}a. The normal irradiation conditions are shown also in Figure~\ref{fig:1}a, where $E=\left(E_{x},0,0\right)\textrm{exp}\left(ikz\right)$ ($H=\left(H_{y},0,0\right)\textrm{exp}\left(ikz\right)$) are the incident electric and magnetic fields, $\vec{k}$ is the wave vector of the normally incident plane wave.

\subsection{\label{sec:sd} Optical properties of single disks}
Since the properties of metasurfaces are determined by the combinations of single particles resonances and the collective resonances of total structure, we start  from a detail discussion the question how the anisotropy of  \ch{MoS2} dielectric permittivity~\cite{NatPhotVolNov}, affects  the scattering of light  by the single \ch{MoS2} disks with the symmetric through hole. In the Cartesian coordinate system shown in Figure~\ref{fig:1}a the tensor of \ch{MoS2} dielectric permittivity $\hat\varepsilon_p$  is written as
\begin{equation}
\label{eq:epsp}
    \hat\varepsilon_p=\left(
    \begin{array}{ccc}
         \varepsilon_\perp&0&0  \\
        0 & \varepsilon_{||}&0\\
        0 & 0& \varepsilon_{||}
    \end{array}
    \right),
\end{equation}
where the values of $\varepsilon_\perp$ and $\varepsilon_{||}$ have been recently measured \cite{NatPhotVolNov,Ermolaev} and shown in Figure~\ref{fig:1}b,c. For calculations of the scattering cross sections and corresponding their multipole decompositions we apply the following approach: the electric and magnetic fields, induced polarization and density of the displacement current inside the scatterers are calculated using the COMSOL facilities for the problem of light plane wave scattering by a finite-size target, and then the  contributions of the main multipole moments into the scattering cross section $\sigma_{scat}$ are determined from the expression \cite{Evlyukhin_PhysRevB_2019}:
\begin{eqnarray}
\label{SCS}
\sigma_{scat}&\simeq&\frac{k^4}{6\pi\varepsilon_0^2|{\bf
E}|^2}|{\bf p}|^2
+\frac{k^4\varepsilon_d\mu_0}{6\pi\varepsilon_0|{\bf
E}|^2}|{\bf m}|^2\nonumber\\
&+&\frac{k^6 \varepsilon_d}{720\pi\varepsilon_0^2 |{\bf E}|^2}\sum_{\alpha\beta}|{Q_{\alpha\beta}}|^2+\frac{k^6
\varepsilon_d^2\mu_0}{80\pi\varepsilon_0 |{\bf
E}|^2}\sum_{\alpha\beta} |{ M_{\alpha\beta}}|^2\nonumber\\
&+&\frac{k^8 \varepsilon_d^2}{1890\pi\varepsilon_0^2 |{\bf E}|^2}\sum_{\alpha\beta\gamma}|{ O_{\alpha\beta\gamma}}|^2,
\end{eqnarray}
where $k$ is the vacuum wave number, $\varepsilon_0$ and $\varepsilon_d$  {are} the vacuum dielectric constant and relative dielectric constant of surrounding medium, respectively, $\mu_0$ is the vacuum permeability, $\bf E$ is the vector  electric amplitude  of the incident plane waves,  $\bf p$ and $\bf m$ are the vectors of electric and magnetic dipole moments of the scatterer, respectively, and $\hat Q$, $\hat M$, and $\hat O$ are the tensors of electric and magnetic quadrupole moments, and electric octupole moment, respectively. Corresponding expressions for  the multipole moments defined by the density of the displacement current can be found elsewhere~\cite{Evlyukhin_PhysRevB_2019}. In this article we consider that the all multipole moments of single nanoparticles is located at their center of mass. This also concerns the calculation of the dipole polarizabilities presented further.

\begin{figure}[t]
\includegraphics[width=0.9\textwidth]{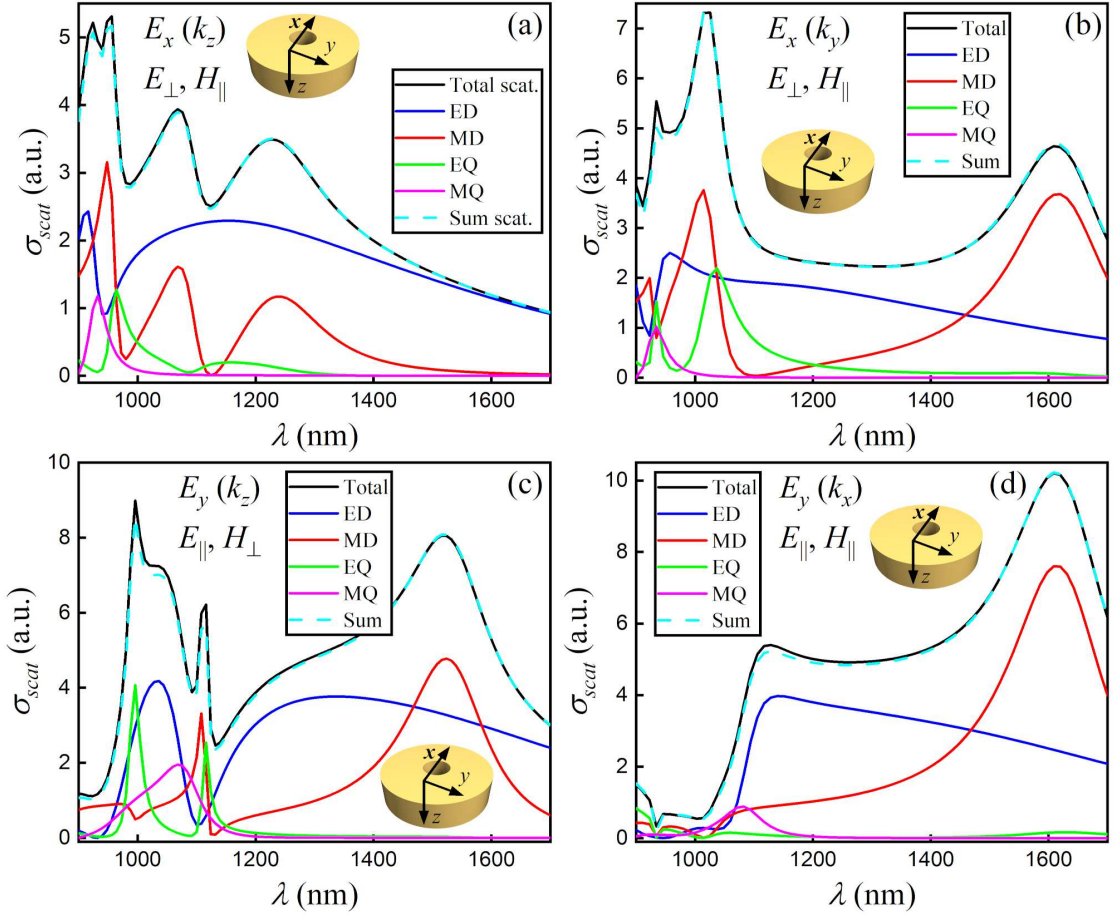}
\caption{\label{fig:SCS} Spectra of the scattering cross sections with corresponding multipole decomposition calculated for the \ch{MoS2} disk (the  radius and height $R_{2}=H=258$ nm)  with the central through hole (the radius $R_{1}=82$ nm) and for different irradiation conditions clarified by the insets. Orientation of the \ch{MoS2} layers is  parallel to $yz$-plane. }
\end{figure}

The scattering cross sections for different irradiation conditions are shown in Figure~\ref{fig:SCS}. Independently on these conditions the scattering in the telecom optical range ($\lambda\in[1250-1650]$~nm) is basically determined by electric dipole (ED) and magnetic dipole (MD) moments of the disk. For normal incidence of the pane waves along $z$-axis (Figure~\ref{fig:SCS}a and~\ref{fig:SCS}c) one can see strong influence of the incident wave polarization on the spectral positions of ED and MD resonances caused by the \ch{MoS2} dielectric anisotropy. Changing the electric polarization from $E_x$ (being perpendicular to \ch{MoS2} layers, Figure~\ref{fig:SCS}a) to $E_y$ (being parallel to \ch{MoS2} layers, Figure~\ref{fig:SCS}c), the value of the cross section increases and the dipole resonances are shifted to the ``red'' side. For the $E_y$-polarization (electric field is directed along the layers), the resonant MD contribution considerably increases. In this case, the MD moment is directed perpendicular to the layers and determined by the ring displacement current induced in these layers characterized by a large refractive index $n\sim\sqrt{\varepsilon_{||}}$. The "red" shift follows from the estimation of the wavelength $\lambda_{res}^{\rm MD}$ corresponding to the MD resonance of  a dielectric particle with dimension $D$ and having refractive index $n$: $\lambda_{res}^{\rm MD}\sim Dn$ -  for fixed $D$, the resonant wavelength increases with $n$.   Thus, the normally propagating light waves with $E_x$-polarization (perpendicular to the layer) do not excite resonant response of the disks around $\lambda=1500$ nm, in contrast to the case of the orthogonal $E_y$-polarization.

For the side irradiation and $H_z$-polarization the results are presented in {Figures}~\ref{fig:SCS}b and~\ref{fig:SCS}d. One can see that the spectral position of the first MD resonance ($\lambda=1620$~nm) does not  {almost} depend on the orientation of the external electric field, whereas the values of the corresponding scattering cross sections are different: for the electric field being parallel  {to} the \ch{MoS2} layers the scattering cross section is larger owing to larger value of the induced displacement current in $xy$-plane. In this case, the value of the resonant MD at $\lambda=1620$~nm is also larger than for other case shown in Figure~\ref{fig:SCS}b. Due to symmetry properties of the disk the MD excited by the side irradiation with $H_z$-polarization has the only $m_z$ component for which
\begin{equation}\label{en_q}
    |m_z^{||}|>|m_z^{\perp}|,
\end{equation}
where $m_z^{||(\perp)}$ is the MD $z$-component for the side irradiation with electric field polarization parallel (perpendicular) to the \ch{MoS2} layers. The inequality (\ref{en_q}) is followed from the anisotropy of $\hat\varepsilon_p$ which also results in the nonlocal optical response of the \ch{MoS2} disk.

In the dipole approximation with inclusion of the fist-order nonlocal response, the electric and magnetic dipole moments of single NPs irradiated by  an external wave with electric $\bf E$ and magnetic $\bf H$ fields can be written in the form \cite{varault2013multipolar}
\begin{eqnarray}
\label{eq:pm}
{\bf p}&=&\hat \alpha^{ee} {\bf E}+\hat\alpha^{em}{\bf H}+\hat a[\nabla\otimes{\bf E}],\\
\label{eq:mm}
{\bf m}&=&\hat\alpha^{mm}{\bf H}+\hat\alpha^{me}{\bf E}+\hat c[\nabla\otimes{\bf E}],
\end{eqnarray}
where $\hat{\alpha}^{ee\left(mm\right)}$ is the tensor of electric (magnetic) dipole polarizability corresponding to the direct (local) excitation of the dipole moment by the electric (magnetic) field of the incident wave. The second and third terms in  {Equations} (\ref{eq:pm}) and (\ref{eq:mm}) correspond to   the bianisotropic and next  nonlocal, respectively,  excitation of $\bf p$ and $\bf m$. From the reciprocity property of the considered scattering systems the tensors of binanisotropic polarizabilityes satisfy the following relation: $\mu_0\hat\alpha^{me}=-(\hat\alpha^{em})^T$ \cite{asadchy2018bianisotropic}, where $T$ denotes the transpose operation. Moreover, from the deriving of (\ref{eq:pm}) and (\ref{eq:mm}) \cite{achouri2021extension}, the elements of the tensors $\hat a$ and $\hat c$ satisfy the following symmetry properties: {$a_{ijl}=a_{ilj}$ and $c_{ijl}=c_{ilj}$}.  Further, we will consider that, when the orientation of the excited dipole is collinear with the wave vector of the incident wave, the bianisotropy is longitudinal, and when it is perpendicular, the bianisotropy is transverse. Note that all tensor coefficients in (\ref{eq:pm}) and (\ref{eq:mm}) depend only {on} geometrical and materials characteristics of scatterers and are independent  on irradiation conditions.

\subsection{Dipole polarizabilities of single nanoparticles}

\begin{figure}
\includegraphics[width=0.75\textwidth]{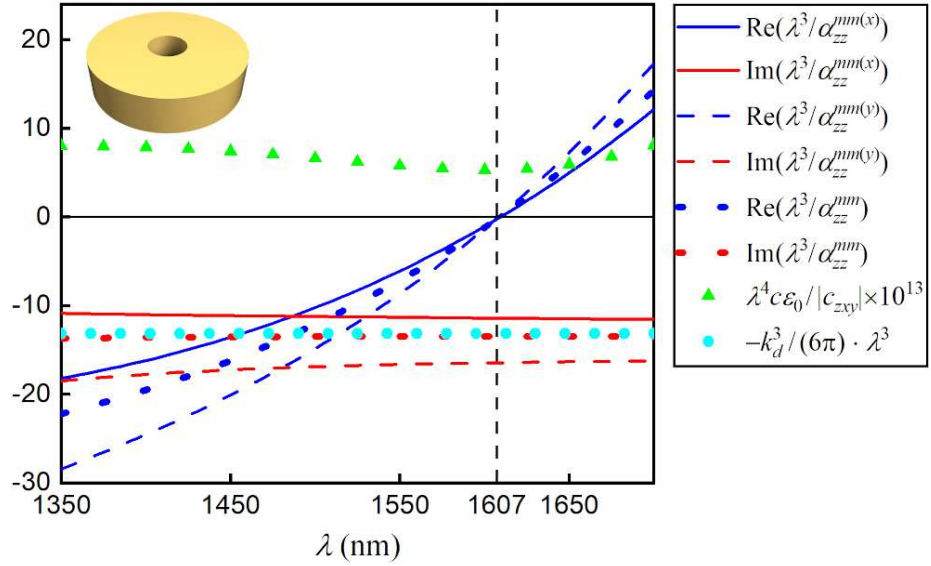}
\caption{\label{fig:3} Real (blue curves) and imaginary (red curves) parts of inverse polarizabilities and nonlocal response (green triangles) calculated according to (\ref{eq:alpmc}) for $\alpha_{zz}^{mm}$ (dotted curves) and derived from equations (\ref{eq:mmc}a)--(\ref{eq:mmc}b) for the effective $\alpha_{zz}^{mm\left(x\right)}$ (solid curves) and $\alpha_{zz}^{mm\left(y\right)}$ (\ref{eq:mmc}c)--(\ref{eq:mmc}d) (dashed curves), separately, for lossless \ch{MoS2} disk with radius and height $R_{2}=H=258 \; \textrm{nm}$ with centered symmetric hole with radius $R_{1}=82 \; \textrm{nm}$ {shown in the inset}. The cyan circles correspond to the analytical expression (\ref{eq:3}b) for the dipole sum.}
\end{figure}

In order to determine  the conditions of  resonant coupling between NPs periodically arranged in  metasurface (as shown in Figure \ref{fig:1}), one needs to know certain components of the tensor coefficients from (\ref{eq:pm}) and (\ref{eq:mm}) which correspond to the lateral irradiation of the single disk. They can be obtained from (\ref{eq:pm}) and (\ref{eq:mm}), if the dipole moments $\bf p$ and $\bf m$ of NP are  known from numerical calculations for the two separate irradiation conditions ({along} $x$- and $y$-direction). For example, in the case of studying coupling effects with a magnetic dipole moments being perpendicular to the metasurface plane,  it is convenient to get from (\ref{eq:pm}) the system of equations in the form:
\begin{subequations}
\label{eq:mmc}
\begin{align}
m_{z}^{\left(k,E,H\right)}&=\alpha_{zz}^{mm}H_{z}+c_{zyx}ikE_{y}+\alpha_{zy}^{me}E_{y},\\
m_{z}^{\left(-k,E,-H\right)}&=-\alpha_{zz}^{mm}H_{z}-c_{zyx}ikE_{y}+\alpha_{zy}^{me}E_{y},\\
m_{z}^{\left(E,k,-H\right)}&=-\alpha_{zz}^{mm}H_{z}+c_{zxy}ikE_{x}+\alpha_{zx}^{me}E_{x},\\
m_{z}^{\left(E,-k,H\right)}&=\alpha_{zz}^{mm}H_{z}-c_{zxy}ikE_{x}+\alpha_{zx}^{me}E_{x}.
\end{align}
\end{subequations}
The Equations (\ref{eq:mmc}a) and (\ref{eq:mmc}b) correspond to the irradiation of NPs in two opposite directions: forward and backward to the $Ox$ axis, while maintaining the direction of the vector \textbf{E} along $Oy$ axis. This information is indicated by the superscripts for $m_z$.  The Equations (\ref{eq:mmc}c) and (\ref{eq:mmc}d) correspond to the irradiation in forward and backward directions along $Oy$ axis while maintaining the direction of the vector \textbf{E} along $Ox$ axis (see the corresponding superscripts for the moment). Using the symmetrical property $c_{zxy}=c_{zyx}$, the solution of system (\ref{eq:mmc})  in the general case, have the form:
\begin{subequations}
\label{eq:alpmc}
\begin{align}
\alpha_{zz}^{mm}&=\frac{1}{4H_{z}}\left[\left(m_{z}^{\left(k,E,H\right)}-m_{z}^{\left(-k,E,-H\right)}\right)
+\left(m_{z}^{\left(E,-k,H\right)}-m_{z}^{\left(E,k,-H\right)}\right)\right],\\
c_{zxy}&=\frac{1}{2ik\left(E_{x}+E_{y}\right)}\left[\left(m_{z}^{\left(k,E,H\right)}-m_{z}^{\left(-k,E,-H\right)}\right)
-\left(m_{z}^{\left(E,-k,H\right)}-m_{z}^{\left(E,k,-H\right)}\right)\right].
\end{align}
\end{subequations}
Note that due to inverse symmetry of the MoS$_2$ disk NP \cite{bobylev2020nonlocal} there is no bianisotropy, but a nonlocal response  can be  excited. This result is a consequence of the anisotropy of NP material since for these irradiation conditions this effect is not observed in isotropic disks \cite{bobylev2020nonlocal}  for which the magnetic dipole component $m_z$ is determined  only by the local term with $\alpha_{zz}^{mm}$ in (\ref{eq:mmc}). Formally, we can also include the nonlocal response in  effective polarizabilities (which become  dependent on irradiation conditions) defining them as 
\begin{equation}
    \alpha_{zz}^{mm(x)}=\frac{m_{z}^{\left(k,E,H\right)}}{H_z }\quad{\rm and}\quad \alpha_{zz}^{mm(y)}=\frac{m_{z}^{\left(E,-k,H\right)}}{H_z }
\end{equation} 
under the irradiation with a wave vector that is collinear to the $Ox$ and $Oy$ axes, respectively. Note that such effective polarizabilites of isotropic disks with a dipole response coincide with their direct polarizabilities $\alpha_{zz}^{mm}$. The inverse values of the corresponding polarizabilites calculated for the disk NP with the central hole as in Figure~\ref{fig:SCS} are presented in Figure~\ref{fig:3} for different  incident wavelengths $\lambda$. The curves corresponding to the direct (local)  dipole polarizability $\alpha_{zz}^{mm}$ and  its effective counterparts $\alpha_{zz}^{mm (x)}$ and $\alpha_{zz}^{mm (y)}$  significantly differ because of  the nonlocal response additions.
Note that the intersection point of the  real parts of the inverse direct (local) and effective  polarizabilities corresponds to their zero values coinciding with the magnetic-dipole resonant condition $\textrm{Re}\left({1}/{\alpha_{zz}^{mm}}\right)=0$  realized at the telecom wavelength $\lambda=1607$~nm, see Figure~\ref{fig:3}. At the same time, the imaginary parts of $1/\alpha_{zz}^{mm(x)}$ and $1/\alpha_{zz}^{mm(y)}$ do not coincide in all considered spectral range including the resonant condition, which leads to the difference in the values of the dipoles $m_{z}^{\left(k,E,H\right)}$ and $m_{z}^{\left(-k,E,-H\right)}$ and, therefore, to different energies of their excitation. 

In the region of small wavelengths around $\lambda=950$ nm  a considerable difference in the scattering cross sections, shown in Figure \ref{fig:SCS}, appears due to anisotropic properties of the disk and high-order multipoles excitation. However, in our further consideration  we shall concentrate  our attention only on  the telecom long-wavelength region where the NP optical properties are determined only by  their dipole response.

\subsection{Quasi-trapped modes of \ch{MoS2} metasurfaces}

\begin{figure}
\includegraphics[width=1\textwidth]{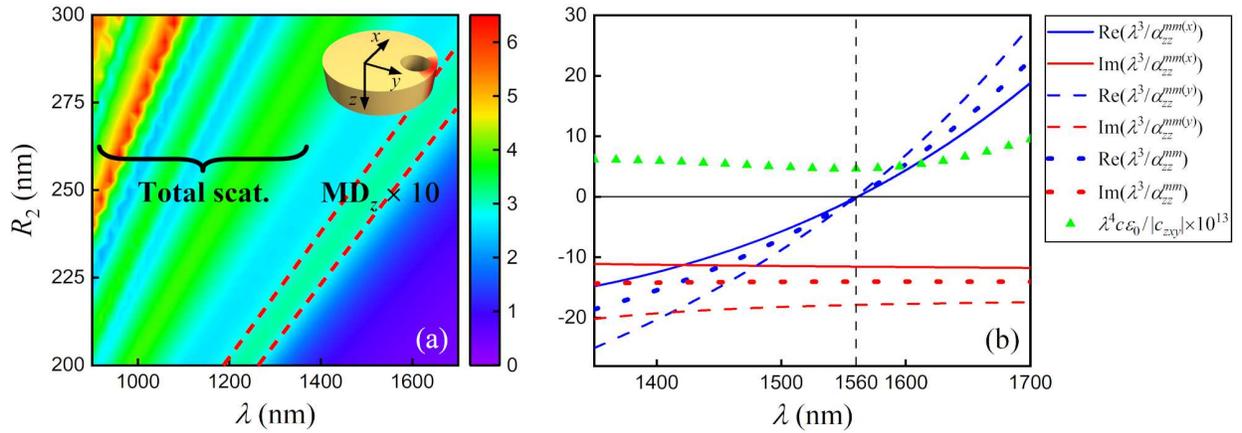}
\caption{\label{fig:4_0} (a) The contour plot on the parametric plane $\left(\lambda; \; R_{2}\right)$ for total scattering cross sections and the partial scattering cross-section associated only with MD component $m_z$ (MD$_z$) calculated for \ch{MoS2} disks  with an eccentric hole normally irradiated by linear $E_x$-polarized waves.   The hole with radius $R_{1}=0.318 R_{2}$ is shifted on $\Delta_{y}=0.38 R_{2}$, {inset in (a) shows the position of the shifted hole in a disk relative to the axes of the Cartesian coordinate system and schematic illustration of hot spot for near field $E_{x}$.{ The disk's height $H=R_2$}. } (b) the same that in Figure~\ref{fig:3}, but for the disk with this shifted hole.}
\end{figure}

According to (\ref{eq:pm}) and (\ref{eq:mm}) under $E=(E_{x},0,0)\textrm{exp}(ikz)$ irradiation conditions, the electric and magnetic dipole moments excited in a single disk NP can be determined as follows:
\begin{subequations}
\label{eq:1}
\begin{align}
p_{i}&=\alpha^{ee}_{ix}E_{x}+\alpha^{em}_{iy}H_{y}+a_{ixz}ikE_{x},\\
m_{i}&=\alpha^{mm}_{iy}H_{y}+\alpha^{me}_{ix}E_{x}+c_{ixz}ikE_{x},
\end{align}
\end{subequations}
In the case $i=z$, second terms correspond to the longitudinal bianisotropy, which can be used for excitation of the quasi-trapped modes of the metasurface composed of such disks, as it has been discussed in Ref. \citenum{B2}. The bianisotropic response of the single disk can appear due to the violation of its inverse symmetry, which is achieved, for example, by shifting a hole on a value $\Delta_{y}$ along the in-plane $Oy$ axis of the Cartesian coordinate system, see inset in Figure~\ref{fig:4_0}a. In this case, the magnetic-type longitudinal bianisotropy, leading to $m_z$, is associated with the term $\alpha_{zx}^{me}E_{x}$ in (\ref{eq:1}). Note that in order to obtain  the longitudinal bianisotropy of the electric type, leading to $p_z$, it is sufficient to change the polarization of the incident wave from $E_x$ to $E_y$.

Spectral positions of multipole resonances of single dielectric disks can be tuned by  variation of their geometrical parameters \cite{evlyukhin2011multipole,staude2013tailoring}. For the MoS$_2$ disks with hole shifted along MoS$_2$ layers with $\varepsilon_{||}$ (along the $y$-axis)  and normally irradiated by plane waves with electric field $E=(E_{x},0,0)\textrm{exp}(ikz)$,   the results of  such tuning procedure are shown in Figure \ref{fig:4_0}a. This figure presents the contour plot of  the scattering cross sections of \ch{MoS2} disks with $H=R_2$, $R_{1}=0.318 R_{2}$ and $\Delta_{y}=0.38 R_{2}$ over the parametric plane $\left(\lambda; \; R_{2}\right)$. For convenience of the result analysis, Figure \ref{fig:4_0}a also includes separate contributions of $m_z$  into the total scattering cross sections excited due to the bianisotropic response with $\alpha_{zx}^{me}$.
One can see that with an increase of the disk radius, its resonances, including the resonant $m_z$ contribution, shift to longer wavelengths. Note that the bianisotropic resonance $ m_z $ has a spectral position separated from the main resonances. The importance of this behaviour will be seen from further consideration. {Note that the redshift of $m_{z}$ resonance is solely due to the orientation of the \ch{MoS2} layers relative to the incident field. In particular, the wavelength of the main scattering resonances for a disk is proportional to its effective permittivity. Since the layers are arranged perpendicular to the exciting field $E_{x}$, this effective permittivity is mainly determined by the component $\varepsilon_{\perp}$. Taking into account that $\varepsilon_{\parallel}>\varepsilon_{\perp}$, the main resonances are in the region of short wavelengths, see Figure~\ref{fig:4_0}a.}

In order to have the bianisotropic $m_z$ resonance in the middle of the telecom range  we choose the disks with $R_2=258$ nm and $H=258$ nm for the formation of metasurfaces. The disks of such  sizes have been studied in Subsection~\ref{sec:sd}, where the dependence of  the spectral positions of their main optical resonances on the irradiation conditions were explained. That obtained results {also qualitatively correspond (excluding the bianisotropy)} to the disks with shifted hole because the shift can be considered as a weak  perturbation.

Before to study  optical properties of metasurfaces composed of the  MoS$_2$ disks with bianisotropic response, let us discuss some general positions related to existence and excitation of  BICs in  {metasurfaces} composed of dipolar particles. As it has been discussed in Refs. \citenum{B2,babicheva2021multipole}, in the case of zero ohmic losses, the collective effect of synchronization of individual magnetic moments $m_{z}$ in an infinite periodic array of holey isotropic disks (with the dipole response)  placed in $xOy$ plane with the period $P$ can lead to the formation of BICs  in the regime of trapped mode  realization. If the magnetic dipole moment $m_z$ is associated only with the direct dipole polarizability $\alpha_{zz}^{mm}$, the condition  of the trapped mode existence corresponds to the  solution of the transcendental equation in the form~\cite{B2,babicheva2021multipole}:
\begin{equation}
\label{eq:2}
S^{\left({\rm R}\right)}_{z}=\textrm{Re}\left(\frac{1}{\alpha^{mm}_{zz}}\right),
\end{equation}
where {$S_{z}^{\left({\rm R}\right)}$} is the real part of the dipole sum $S_{z}=S^{\left({\rm R}\right)}_{z}+iS^{\left({\rm I}\right)}_{z}$ with infinite number of terms. Its parts can be determined as follows:
\begin{subequations}
\label{eq:3}
\begin{align}
S^{\left({\rm R}\right)}_{z}&=\frac{k^{2}_{d}}{4 \pi}\sum_{l,j}\left(\frac{\textrm{cos}\left(k_{d}d_{lj}\right)}{d_{lj}}
-\frac{\textrm{sin}\left(k_{d}d_{lj}\right)}{k_{d}d^{2}_{lj}}-\frac{\textrm{cos}\left(k_{d}d_{lj}\right)}{k^{2}_{d}d^{3}_{lj}}\right),\\
S^{\left({\rm I}\right)}_{z}&= -\frac{k^{3}_{d}}{6\pi},
\end{align}
\end{subequations}
where the last equality is obtained in  the non-diffractive  regime for which the period $P$ of the array is smaller than the incident  {wavelength} so that $k_dP<2\pi$.
Here, the parameter $d_{lj}=P\sqrt{l^{2}+j^{2}}$ is the distance from the Cartesian coordinate system origin, coinciding with position of a lattice node, to all other nodes of the lattice which are numbering by indices $l$ and $j$, $k_{d}=2\pi /\lambda_d$, and $\lambda_{d}$ is the incident  wavelength in the surrounding medium with relative dielectric constant $\varepsilon_d$. In this paper we consider that $\varepsilon_d=1$.

Using (\ref{eq:2}) and (\ref{eq:3}a) together with $\alpha_{zz}^{mm}$, it is possible to determine the lattice period $P$ for a given wavelength of trapped mode $\lambda_{TM}$ \cite{Evlyukhin_Arxiv2021}. For example, we can fix the single particle  polarizability $\alpha_{zz}^{mm}$ at a certain wavelength ($\lambda_{TM}$) and then, solving (\ref{eq:2}), to find the period $P$ of a metasurface (with a square elementary cell) that will support the trapped mode state at this $\lambda_{TM}$. Note, since the imaginary part $S^{(I)}$ of the dipole sum in (\ref{eq:3}b) does not depend on the lattice period $P$, the  Equation (\ref{eq:3}b) is always correct in the non-diffractive regime. Let's consider the equation 
\begin{equation}\label{Im}
S^{\left(I\right)}_{z}=\textrm{Im}\left(\frac{1}{\alpha^{mm}_{zz}}\right),
\end{equation}
providing together with  {Equation} (\ref{eq:2}) the second condition for the existence of  a symmetry protected bound state in the continuum or, by other words, a true trapped mode with infinite quality factor. However, the condition (\ref{Im}) can be  exact satisfied  only in the case of  the dipole response without ohmic losses,  when  ${\rm Im}(1/\alpha_{zz}^{mm})=-k_d^3/(6\pi)$ \cite{B2,babicheva2021multipole}.

In general, the electric and magnetic dipole moments of NPs can include the contributions stemming on the nonlocalyty including the  bianisotropy, see  {Equations} (\ref{eq:pm}) and (\ref{eq:mm}), and the absorption. In these cases, the metasurfaces composed of such dipole NPs cannot support the ideal BICs, but there can be realized {so-called q-BICs or QTMs} characterized by finite   quality factors, the values of which are determined by the bianisotropy contributions. However, as shown our calculations, even in the cases of the nonlocality, the relation (\ref{eq:2}) only with  $\alpha_{zz}^{mm}$, can be used for  determination  of the lattice period $P$ of the  metasurface  supporting a QTM at  $\lambda_{TM}$. We assume that this follows from the fact that only dipole-dipole interaction  is involved in the QTM formation, while a nonlocal response  can be related with the generation of scatterer dipole moments  by high order multipole parts of the incident wave~\cite{Varault2013-Multipolareffectson}. 

Now we examine a lossless \ch{MoS2} disk with an eccentric hole, i.e., with the imaginary parts  $\varepsilon_{\perp}^{\left(\textrm{I}\right)}=\varepsilon_{\parallel}^{\left(\textrm{I}\right)}=0$ and with the bianisotropic response. As it was discussed above, we chose $H=R_2$, $R_{2}=258$~nm, $R_{1}=82$~nm, $\Delta_{y}=98$~nm and calculate the polarizabilities  $\alpha_{zz}^{mm}$, and $\alpha_{zz}^{mm(x)}$, $\alpha_{zz}^{mm(y)}$ of disk according with (\ref{eq:alpmc}a). The forms of the curves for the real and imaginary parts of  $1/\alpha_{zz}^{mm}$, and $1/\alpha_{zz}^{mm(x)}$, and $1/\alpha_{zz}^{mm(y)}$ shown in Figure~\ref{fig:4_0}b  remain qualitatively same as in Figure~\ref{fig:3}. However, due to the  bianisotropy induced by the hole shift, the resonance condition ${\rm Re}\left({1}/{\alpha_{zz}^{mm}}\right)=0$ is realized at new wavelength $\lambda\approx1562$~nm (Figure~\ref{fig:4_0}b). Note that ${\rm Re}\left({1}/{\alpha_{zz}^{mm(x)}}\right)=0$ and ${\rm Re}\left({1}/{\alpha_{zz}^{mm(y)}}\right)=0$ are also realized at this wavelength.

Thus, the considered anisotropic disk with an eccentric hole demonstrates the bianisotropic response   in the IR range. It can be used as a building block  for creation of a metasurface supporting  a QTM in a given spectral range. {Choosing the wavelength}  $\lambda_{QTM}=1552$ nm{, at which a QTM is assumed to be realized}, and taking ${\rm Re}(1/\alpha_{zz}^{mm})$ at this wavelength ({Figure}~\ref{fig:4_0}b) we obtain from the solution of  {Equation} (\ref{eq:2}) the period {$P=1063 \; {\rm nm}$} of the metasurafce.  {Note that the chosen combination of the wavelength of the incident wave and the cylinder's size correspods to the region of strong bianisotropic response $m_{z}$ in Figure~\ref{fig:4_0}a. This situation} should guarantee a finite width of the QTM resonance because, in this case, ${\rm Im}(S_{z})>{\rm Im}({1}/{\alpha_{zz}^{mm}})$.

\begin{figure}[t]
\centering
\includegraphics[width=0.8\columnwidth]{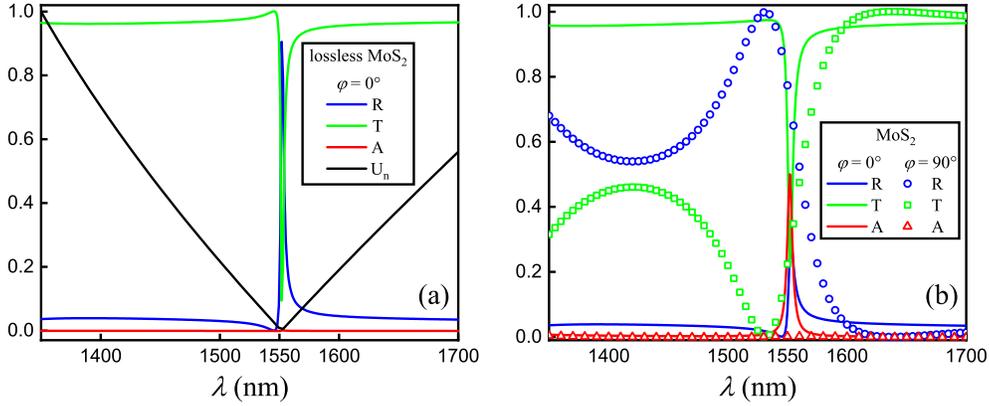}
\caption{\label{fig:4} Spectra of the  reflection (R), transmission (T), absorption (A) coefficients and parameter $U_{n}$ for array with period $P=1063$ nm composed of (a) ideal lossless and (b) real weakly-absorbing \ch{MoS2} disks; $U_{n}=U/{\rm max}\left(U\right)$, where $U=\left|S_{z}^{\left({\rm R}\right)}-{\rm Re}\left(1/\alpha_{zz}^{mm}\right)\right|$; the condition $U_{n}=0$ indicates the position of QTM. The geometric disk parameters correspond to Figure~\ref{fig:4_0}b. The angle $\varphi$ determines the electric polarization of the incident waves: $\varphi=0^{\rm o}$ ($\varphi=90^{\rm o}$) determines the $E_x$($E_y$)-polarization.  }
\end{figure}

For the verification of the   dipole approximation we numerically calculated, using COMSOL, the transmission and reflection spectra of the metasurface ({Figure}~\ref{fig:1}a) composed of the hollow disks with the size parameters as in  {Figure}~\ref{fig:4_0} and with the period $P\approx 1063$~nm. As it is demonstrated in {Figure}~\ref{fig:4}a for the lossless case, the transmission and reflection spectra have  narrow resonant  {features} just at the wavelength 1552 nm of the QTM. 
As it is followed from the dipole approxinmation \cite{B2}, at this spectral {region} the strong resonant $m_z$ component of the  {every} disk  is  excited due to the realization of the QTM conditions. Simultaneously, owing to {bianisotropy} there is also connection $m_z$ with $p_x$ \cite{B2}, so that the electric dipole component $p_x$ of the every disk  is also resonantly increased  resulting in suppression of the transmission and increasing of reflection. Thus this behaviour is fully agreed with considerations provided by the dipole approximation. Note that the resonant values of $m_z$ under QTM conditions lead to an increase in the near field in the metasurface plane, since disks do not emit electromagnetic waves perpendicular to this plane \cite{B2}.

\begin{figure}[t]
\centering
\includegraphics[width=1.\columnwidth]{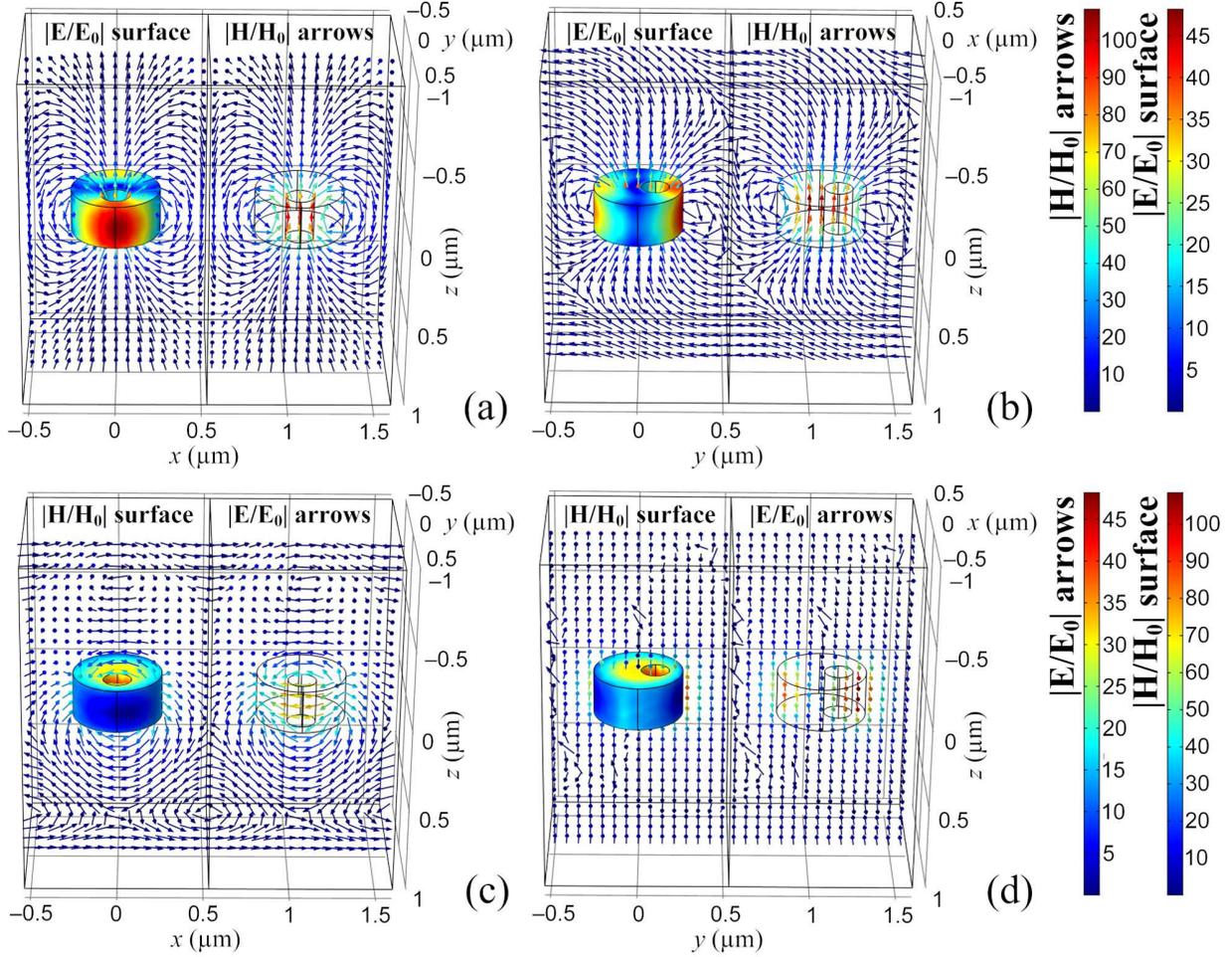}
\caption{\label{fig:5} The visualization of electric and magnetic field distributions calculated  (a), (c) in the $xz$-plane and (b), (d) in the $yz$-plane for  a pair of \ch{MoS2} disks in the metasurface at the conditions of the QTM. The metasurface parameters correspond to Figure~\ref{fig:4}a.}
\end{figure}

Further, we will take into account that MoS$_2$ has small but finite values of the imaginary parts of its dielectric  {permittivities} in the  telecommunication range ({Figure}~\ref{fig:1}d). As a result the absorption effect has to be realized in the metasurface  with parameters as in {Figure}~\ref{fig:4}a and with actual MoS$_2$ dielectric permittivity.  Results of the transmission, reflection and absorption spectra simulations for the two different polarization of the incident waves are shown in Figures. \ref{fig:4}b. Remarkably, that for the $E_x$-polarization, the {joint occurrence of} dissipative effects in each building block and the collective effects of the QTM formation in the entire metasurface leads to the appearance of a narrow peak of collective losses at the wavelength of the QTM, see Figure~\ref{fig:4}b. {This is due to the considerable concentration of the electric field inside the cylinders under the QTM conditions. Indeed, the coupling between individual cylinders in the array occurs due to the interaction of their magnetic fields. They are significantly enhanced in the holes and induce the electric field hot spots inside the cylinders, which are the centers of collective absorption, see Figure~\ref{fig:5}.} This allows to use such a dissipative metasurface for ultra-narrowband absorbing in IR range or for highly sensitive detection of nanoobjects near it. Analyzing the absorption{magenta}{properties} of the system, we found that under the selected conditions in Figure~\ref{fig:4}b, the absorption efficiency is  high and equals to $51\%$ at $\lambda_{QTM}=1552 \; \textrm{nm}$. The quality factor {$Q=\lambda/\textrm{FWHM}$} is $310$ for the {FWHM} of absorption peak equals only to $5 \; \textrm{nm}$. In addition, the system demonstrates a high $S=380 \; \textrm{nm}/\textrm{RIU}$ sensitivity to the environment and reaches the level of figure of merit (FOM) $\textrm{FOM}=76$, compare with Refs.~\citenum{ABS1,ABS2,OSA1,Wang4}. Here we used the relations $S=\Delta \lambda/\Delta n$ and $\textrm{FOM}=S/\textrm{FWHM}$, where $\Delta \lambda$ is the wavelength shift of reflectance resonance as result of the environment refractive index $\Delta n$ variation. 

Another feature of the metasurface composed of the bianisotropic MoS$_2$ disks is a high polarization sensitivity. When the electric field polarization  of the incident wave is rotated by an angle $\varphi =\pm90^{\rm o}$ the QTM, including the collective dissipative effect, completely disappears after the vanishing of the longitudinal bianisotropy of the magnetic type, see the transmission, reflection, and absorption spectra in {Figure~\ref{fig:4}b} calculated for $\varphi=90^{\rm o}$. Here  $\varphi$ is the rotation angle in the $xy$-plane of the electric polarization  relative to the $x$-axis: $\varphi=0^{\rm o}$ ($\varphi=90^{\rm o}$) corresponds to the $E_x$($E_y$)-polarization.

\begin{figure}[t]
\centering
\includegraphics[width=0.8\columnwidth]{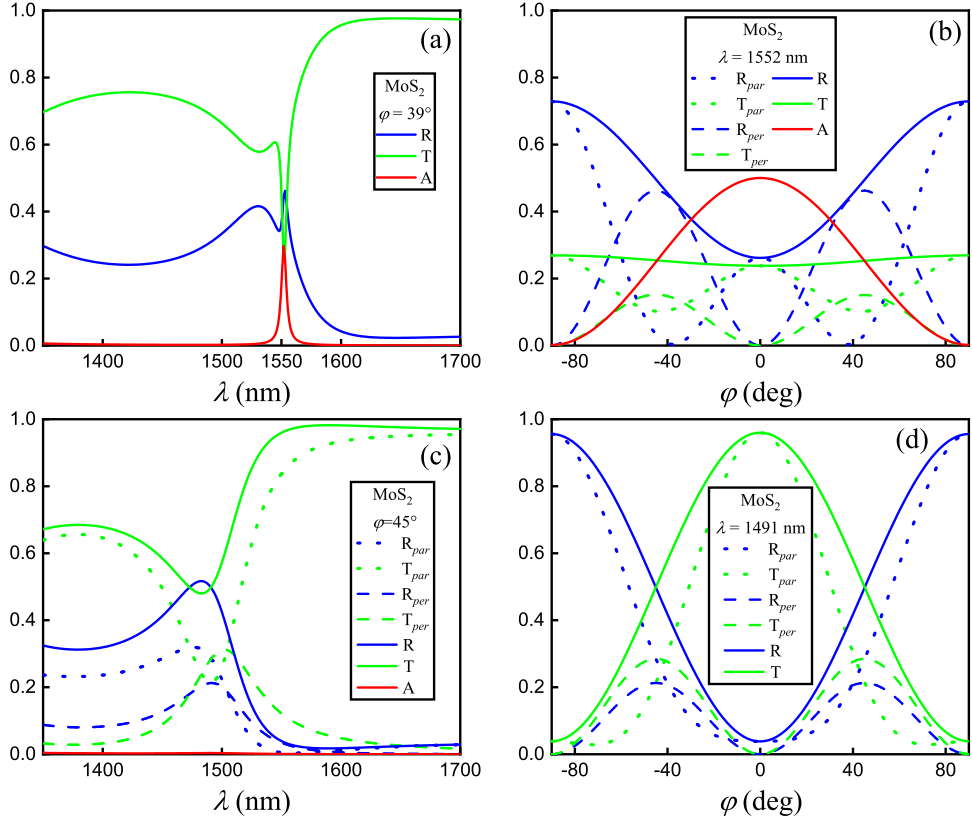}
\caption{\label{fig:4_1} Spectral and angular dependencies for the total reflection (R), transmission (T) and absorption (A) coefficients and their collinear (${\rm R}_{par}$, ${\rm T}_{par}$) and orthogonal (${\rm R}_{per}$, ${\rm T}_{per}$) polarization components of the incident wave for array with period $P=1063$~nm of  absorbing \ch{MoS2} disks with (a), (b) shifted hole and parameters corresponding to Figure~\ref{fig:4_0}b and with (c), (d) centered hole and parameters corresponding to Figure~\ref{fig:3}. The angle $\varphi$ determines orientation of the electric polarization of the incident waves with respect to $x$-axis. }
\end{figure}

In addition, we investigated the polarization sensitivity over the entire range of angle $\varphi$ at the wavelength {$\lambda_{QTM}=1552 \; \textrm{nm}$} and found the decreasing of the absorption maximum associated with the QTM with changing the angle $\varphi$ from $0^{\rm o}$ to $\pm 90^{\rm o}$. For the case $\varphi=39^{\circ}$ the result is shown in Figure~\ref{fig:4_1}a. Note that the spectral position of the absorption maximum does not depend on the angle $\varphi$ always remaining at the wavelength $\lambda_{QTM}=1552 \; {\rm nm}$. The dependence of the transmission, reflection, and absorption coefficients on the angle $\varphi$ at this wavelength is shown in Figure~\ref{fig:4_1}b. From this figure, we observe another feature: the rotation of the polarization of the normally incident wave initiates the process of transferring part of its energy into the components of the transmitted and reflected waves, which are orthogonal to the polarization of the incident wave. In particular, choosing the angle $\varphi=39^{\circ}$ leads to the generation of the reflected wave only with linear polarization that is orthogonal to the polarization of the incident wave, see curve ${\rm R}_{per}$ in Figure~\ref{fig:4_1}b. The maximum of the energy transformation into orthogonal component simultaneously the transmitted and reflected waves is realized at the  angle $\varphi=45^{\circ}$ for both parameters ${\rm R}_{per}$ and ${\rm T}_{per}$ (Figure~\ref{fig:4_1}b). 

{In order to exclude any influence of the particle bianisitropy on the cross-polarization effect in the transmission and reflection, we analyzed this effect  for the metasurface composed of disks with centered holes and also found its presence in the system, see Figure~\ref{fig:4_1}c,d. However, in this case, the maximum for ${\rm R}_{per}$ is realized at the wavelength $1491$~nm (see Figure~\ref{fig:4_1}c). Figure~\ref{fig:4_1}d, plotted for this wavelength,   demonstrates the conservation of the angular position of the maximum for ${\rm R}_{per}$ (${\rm T}_{per}$) at $\varphi=45^{\circ}$. Our simulations of the metasurfaces, composed of the disks with similar geometry fabricated from isotropic material, showed  that such metasurfaces does not possess the property of polarization transformation. Thus, we can conclude that  the observed effects are associated with the anisotropic properties of single building blocks. Hence,  metasurfaces of subwavelength thickness composed of single in-plane symmetric  building blocks of anisotropic materials can have properties of a solid plate consisting of birefringent material.}

In the final part of the article we note another important feature of the quasi-trapped modes in the metasurface of anisotropic NPs. This is the presence of the preferred  direction of the interaction between them. {Figure~\ref{fig:5} shows the distribution patterns of the electric (by colormap on the cylinder surface in Figures~a,b and by vectors in the chosen plane in Figures~c,d) and magnetic (by vectors in the chosen plane in Figures~a.b and by colormap on the cylinder surface in Figures~c,d) fields for the disks in metasurface plotted for different views. It can be seen that a strong coupling between single disks is realized by connecting the magnetic field lines in the $zx$-plane, see Figure~\ref{fig:5}a. At the same time, the magnetic field is tighten and concentrated in the holes of the disks with the formation of hot spots of the electric field in the narrow waist near them. On the contrary, the visualization of the magnetic field lines in the $yz$-plane demonstrates their distancing or ``anticouplin'', see Figure~\ref{fig:5}b. A similar character of coupling/anticoupling is observed for the vectors of electric field in the space between the cylinders in Figure~\ref{fig:5}c,d.} We figured out that such an asymmetric field distribution does not depend on the way of same QTM excitation, but is {a particular feature of the using anisotropic materials for fabricating metasurfaces}. In particular, the same character of the coupling is kept in the case of shifting the hole along $x$-axis and using the field $E=\left(0,E_{y},0\right)\textrm{exp}\left(ikz\right)$ for irradiation of array of disks with holes.

\section{Conclusion}
In present work, we have studied the features of the optical response of disk NPs fabricated from an optically anisotropic material MoS$_2$. The scattering cross sections for different irradiation directions have been calculated. It has been shown that spectral positions of multipole resonances are determined not only the shape and size of the scatterer but also the internal material anisotropy which can affect nonlocal optical properties of the scatterers.   We have investigated the features of transmission and reflection spectra  of metasurafces composed of such nanoparticles and determined conditions for the excitation of QTM leading to formation of a narrow absorption  band in the telecom spectral range. It has been also found the effect of the polarization transformation leading to the appearance of reflection and transmission waves with the orthogonal  polarization with respect to the polarization of the incident wave.  The important role of the material anisotropy in these transformations has been demonstrated. The obtained results can be used  for the design of all-dielectric {metasurfaces} intended for narrow-band quenching of IR signals and fabrication of polarization-sensitive distributed sensors {and polarization metacoatings}.

\begin{acknowledgement}

This work was supported by the Russian Science Foundation, Grant No. 20-12-00343.

\end{acknowledgement}

\bibliography{NL}

\end{document}